\documentclass[a4paper]{jpconf}
\usepackage{graphicx}
\graphicspath{{img/}}
\usepackage{ amssymb }
\begin{document}
\title{Towards Reliable Neural Generative Modeling of Detectors}

\author{L~Anderlini$^{1}$, M~Barbetti$^{1,2}$, D~Derkach$^3$, N~Kazeev$^3$, A~Maevskiy$^3$, S~Mokhnenko$^3$\newline on behalf of LHCb collaboration}

\address{$^1$ Istituto Nazionale di Fisica Nucleare - Sezione di Firenze, via G. Sansone, 1, Sesto Fiorentino, Italy}
\address{$^2$ Dipartimento di Ingegneria dell'Informazione, Università degli Studi di Firenze, via Santa Marta, 3, Firenze, Italy}
\address{$^3$ HSE University, 20 Myasnitskaya st., Moscow 101000, Russia
}

\ead{smohnenko@hse.ru}

\begin{abstract}
The increasing luminosities of future data taking at Large Hadron Collider and next generation collider experiments require an unprecedented amount of simulated events to be produced. Such large scale productions demand a significant amount of valuable computing resources. This brings a demand to use new approaches to event generation and simulation of detector responses. In this paper,
we discuss the application of generative adversarial networks (GANs) to the simulation of the LHCb experiment events. We emphasize main pitfalls in the application of GANs and study the systematic effects in detail. The presented results are based on the Geant4 simulation of the LHCb Cherenkov detector.
\end{abstract}

\section{Introduction}
\label{sec:intro}
At the moment, the Large Hadron Collider (LHC) is preparing for the data-taking period, Run 3, for which it is planned to increase the luminosity for LHCb experiment~\cite{Fartoukh:2790409}. The order of magnitude increase in luminosity will require a large number of simulated events to perform physics analyses. Since pledges on computing resources don't scale as fast as luminosity, traditional detector simulation techniques based on Monte Carlo methods (MC) modelling the radiation-matter interactions~\cite{Ferrari:898301, Agostinelli:2002hh} must be complemented and partially replaced with Fast Simulation options.
An interesting alternative to detailed simulation are parametric simulations, where the relationship between incident particle kinematics and observables is obtained using physics-motivated relations. With one of the possible universal approximators being neural networks, machine-learning driven simulation methods~\cite{Paganini:2017hrr} are getting more and more popular in high-energy physics experiments~\cite{Maevskiy:2020ank,Chekalina:2018hxi,ATL-SOFT-PUB-2020-006,Musella:2018rdi,Fanelli:2019qaq}.

The LHC includes four main experiments: ALICE, ATLAS, CMS, and LHCb. The latter, LHCb, is a single-arm forward spectrometer originally conceived for studies on CP-symmetry violations and rare decays in the $b$-sector.
For the purpose of many of these measurements, LHCb is equipped with two Ring Cherenkov Detectors (RICH) optimized to allow an excellent kaon-pion separation within a wide range of momentum and pseudorapidity and to provide, in general, outstanding Particle Identification (PID) performance~\cite{LHCbRICHGroup:2012mgd}.
Full simulation of the RICH detectors is one of the most CPU-expensive step in the LHCb simulation since it requires accurate modeling of optical photon propagation with diffraction and absorption effects, as well as processes with low-energy secondary electrons~\cite{Easo:2005xv}.

In this article, we discuss a GAN-based approach for an ultra-fast machine-learning driven simulation of the RICH sub-detectors of the LHCb experiment. We continue developing the previously proposed approach~\cite{Maevskiy_2020} for training GANs on real data, with the main emphasis on evaluating the systematic effects arising due to effective neural-network based parameterization. This is done using LHCb simulated samples.

\section{RICH detector and its data}

The principle of operation of the RICH sub-detector is based on the Cherenkov effect. 
A particle moving through the medium at a velocity higher than the phase velocity of light in the medium emits Cherenkov photons. 
The photons are emitted in a cone whose spread angle is a function of the particle’s velocity.
Measuring this angle, via the radius of a reflected ring, and knowing its momentum, allows to identify the particle by constraining its mass.  

The quantities obtained from the RICH reconstruction algorithm are \texttt{RichDLLx} where \texttt{x} can denote kaons - \texttt{k}, protons - \texttt{p}, muons - \texttt{mu}, electrons - \texttt{e} and below threshold - \texttt{bt}. 
\texttt{RichDLLx} for each track is defined in terms of the difference between the logarithmic likelihood for a given particle type hypothesis and the pion hypothesis for that track~\cite{Forty:1998eqa} as
$$
\mathtt{RichDLLx} = \log \mathcal L \left(t_i = \mathrm x, \{t_j\}_{j\neq i} = \{\hat t_j\}_{j\neq i} \right) - \log \mathcal L \left(t_i = \pi,\{t_j\}_{j\neq i} = \{\hat t_j\}_{j\neq i} \right)
$$
where\\
$\mathcal{L}\left(t_1,... , t_N \right)$ – likelihood to observe a given picture, as a function of all charged particle types, \\
$t_i$ – hypothesized particle type for track $i$, \\
$\pi$ -- a pion hypothesis, \\
$\left( \hat t_{1}, ..., \hat t_{N}\right)$ -- a hypothesis maximizing $\mathcal{L}$ is searched for. 

The task of the data-driven approach to modeling the RICH detector is to learn how to simulate the distributions of \texttt{RichDLLx} values for different decays of interest for physics measurements. 
In our previous work~\cite{Maevskiy_2020}, we show that fast simulation of the RICH detectors can be accurately performed using GANs~\cite{NIPS2014_5ca3e9b1}. 
The proposed model shows good approximation to the real data distributions. 
This approach allows to speed-up the simulations production with respect to with detailed simulation, and, in addition, being trained using real data it is not affected by the intrinsic bias of simulation. However, it requires a study of effects arising due to effective GAN parameterization.

In this article, we explore whether our model allows us to control systematic uncertainties in a real physics analysis scenario.
We study how well our model generalizes to the decays not seen during training. For this we use MC samples. 
We use the track reconstructed data as input conditional variables: momentum ($P$), pseudorapidity ($\eta$) and number of hits in the Scintillating Pad Detector (nSPDHits)~\cite{PicatosteOlloqui:2009xwb}.

\section{Our model architecture}
Generative Adversarial Networks (GAN) is a powerful class of generative models based on the simultaneous training of two neural network. The first network, called generator, generates synthetic samples, while the other one, called discriminator, tries to distinguish real samples from those produced by the generator. These two networks learn to compete with each other in a zero-sum game. In this way, the generator learns to produce samples that do not differ from the real ones.

As a starting point for our model, we use the Cramér-GAN~\cite{DBLP:journals/corr/BellemareDDMLHM17}. This GAN flavor uses a metric  between distributions, called the Energy distance (multivariate generalization of the Cramér distance). It preserves all the nice properties of the Wasserstein GAN~\cite{arjovsky2017wasserstein}, while solving the biased  gradients problem~\cite{DBLP:journals/corr/BellemareDDMLHM17}.

\begin{figure}
    \centering
    \includegraphics[width=0.5\textwidth]{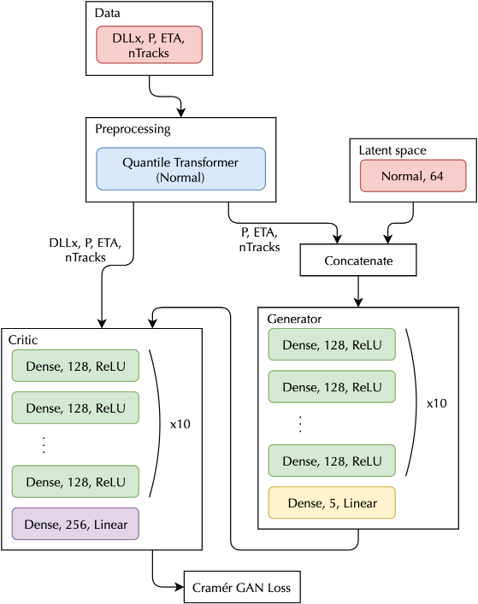}
    \caption{The architecture of our model}
    \label{fig:network}
\end{figure}

The architecture of our neural network is shown in Figure~\ref{fig:network}. This architecture is demonstrated to be sufficient~\cite{Maevskiy_2020} to describe the RICH variables with high accuracy when trained on the reconstructed calibration samples from the real data obtained with the LHCb detector~\cite{Aaij:2018vrk}. Calibration samples are special datasets selected and reconstructed avoiding selection bias on a set of probe particle species. With a novel data-driven training based on the sWeights background subtraction~\cite{sweight2005, sweight2019} the quality of the description obtained is sufficiently high.

\section{Results}
The purpose of this study is to show the transferability of our models to decays not present in the training set. In order to work with clean decay signatures, the study is performed on the detailed MC samples. We want to emphasize that while this procedure is performed using simulated samples, globally, our model is designed to be trained using real data samples. In this paper, we show our results of training the GAN on muons from a mixture of simulated events: inclusive $J/\psi$ and $B^{\pm} \to J/\psi(\mu^{+}\mu^{-})K^{\pm}$ and evaluating this GAN on the $B^{\pm} \to K^{*\pm}\mu^{+}\mu^{-}$ test decay. 

\begin{figure}
    \centering
    \includegraphics[width=0.9\textwidth]{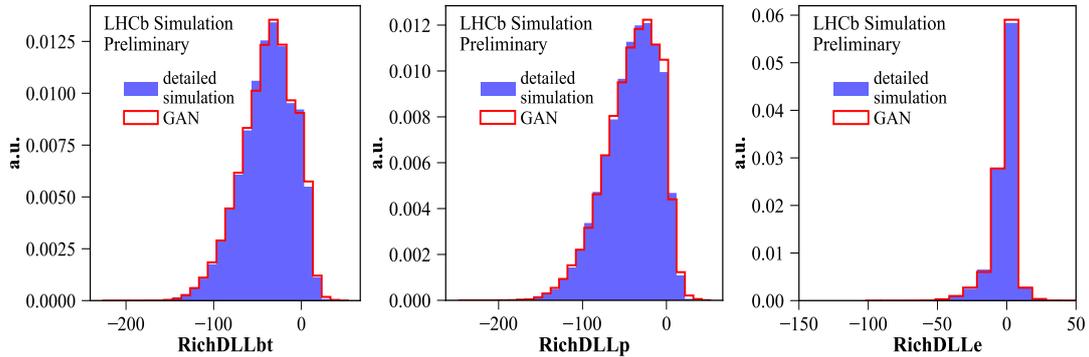}
    \caption{Histograms of distributions of real and generated output variables for the test decay $B^{\pm} \to K^{*\pm}\mu^{+}\mu^{-}$.}
    \label{fig:DLLDistributuions}
\end{figure}

Figure~\ref{fig:DLLDistributuions} shows the \texttt{RichDLLx} variable distributions for the detailed simulation and data generated by our GAN on the test decay channel. 
These results show that the GAN performance remains stable despite the change of training set (from real data sample to simulated one). However, since not only the global distribution match is important in physics analysis, we also study the quality of the description as a function of the input parameters. To do that, we introduce the efficiency ratio metric as follows. 
\begin{enumerate}
  \item Measure the efficiency of \texttt{RichDLLx} cuts at various quantiles of the \texttt{RichDLLx} distribution:
    $$\epsilon=(number\:of\:tracks\:above\: threshold)/total\:number\:of\:tracks.$$
  \item Do this as a function of the input variables: $\epsilon(P, \eta, nSPDHits)$.
  \item Calculate the efficiency ratio between GAN predictions and simulated events 
(in bins of a variable): 
    $$\mathit{efficiency}\:ratio= \epsilon_{GAN}/\epsilon_{simulated}$$
\end{enumerate}

\begin{figure}
    \centering
    \includegraphics[width=0.9\textwidth]{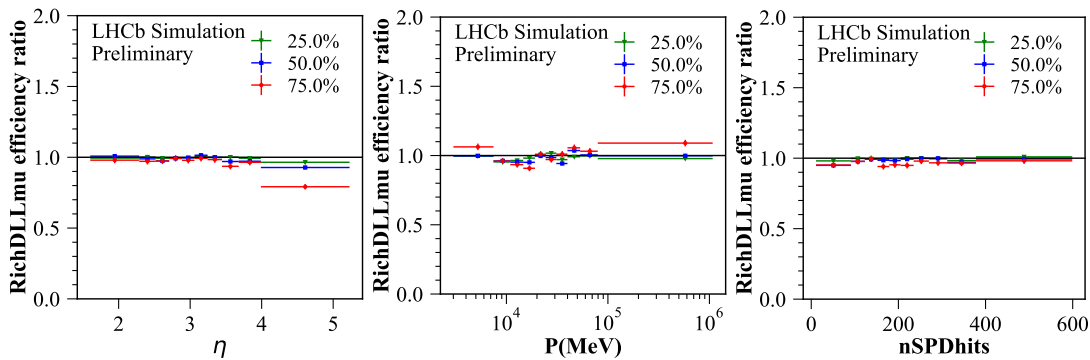}
    \caption{Dependence of \texttt{RichDLLmu} efficiency ratio on input variables: 
    momentum (P), 
    pseudorapidity ($\eta$), 
    number of hits in the Scintillating Pad Detector (nSPDHits) for the test decays $B^{\pm} \to K^{*\pm}\mu^{+}\mu^{-}$ unseen by GAN during training.}
    \label{fig:mu}
\end{figure}

Figure~\ref{fig:mu} shows dependence of \texttt{RichDLLmu} efficiency ratio between GAN predictions and detailed simulated events on input variables for three quartiles. In case of an ideal trained GAN this ratio should be close to 1, thus showing that the efficiencies can be predicted precisely using GAN model. 
We have good agreement between simulation and data in three projections, although quality of agreement degrades in the tails of the distributions.

\begin{figure}
    \centering
    \includegraphics[width=0.45\textwidth]{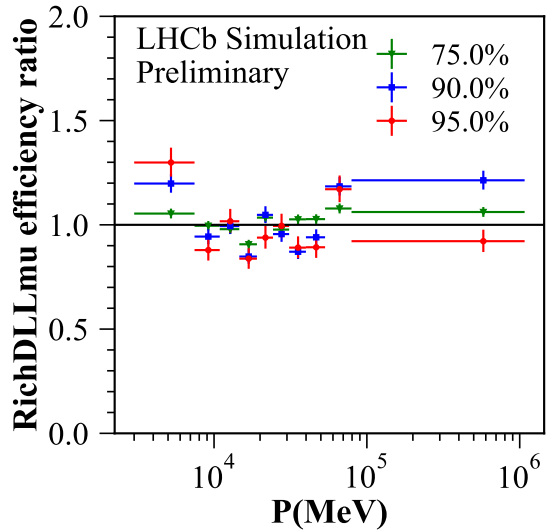}
    \caption{Dependence of \texttt{RichDLLmu} efficiency ratio on momentum (P).}
    \label{fig:RichDLLmu}
\end{figure}

Figure~\ref{fig:RichDLLmu} shows the dependence of \texttt{RichDLLmu} efficiency ratio on the momentum (\texttt{P}) with 75, 90, and 95\% selection efficiencies. In this region, one can see that the quality of the description degrades on the tails of the distributions. 
At low momenta, the difference can be up to 50\%. While this problem is quite significant, we do would like to stress that we are talking about the tails of the distribution, thus the overall description is not affected significantly.  
We expect the problem to be less pronounced as learning statistics and model complexity increase.

\section{Conclusion}
\label{sec:conclusion}
We show that using a GAN-based approach for fast simulation of RICH sub-detectors in LHCb provides good description of the efficiencies. Some effects observed in the tails of the distribution do not affect the overall conclusion. After testing the quality on the decays unseen during training, we conclude that the description transfer is also robust and thus can be used for real-life physics analysis in LHCb.

\section*{Acknowledgement}
The research leading to these results has received funding from Russian Science Foundation under grant agreement 17-72-20127. This research was supported in part through computational resources of HPC facilities at HSE University~\cite{Kostenetskiy_2021}.

\section*{References}
\bibliographystyle{iopart-num}
\bibliography{refs}

\providecommand{\newblock}{}
\begin{thebibliography}{10}
\expandafter\ifx\csname url\endcsname\relax
  \def\url#1{{\tt #1}}\fi
\expandafter\ifx\csname urlprefix\endcsname\relax\def\urlprefix{URL }\fi
\providecommand{\eprint}[2][]{\url{#2}}

\bibitem{Fartoukh:2790409}
Fartoukh S, Kostoglou S, Solfaroli~Camillocci M, Arduini G, Bartosik H, Bracco
  C, Brodzinski K, Bruce R, Buffat X, Calviani M, Cerutti F, Efthymiopoulos I,
  Goddard B, Iadarola G, Karastathis N, Lechner A, Metral E, Mounet N, Nuiry
  F~X, Papadopoulou P~S, Papaphilippou Y, Petersen B, Persson T~H~B, Redaelli
  S, Rumolo G, Salvant B, Sterbini G, Timko H, Tomas~Garcia R and Wenninger J
  2021 {LHC Configuration and Operational Scenario for Run 3} Tech. rep. CERN
  Geneva \urlprefix\url{https://cds.cern.ch/record/2790409}

\bibitem{Ferrari:898301}
Ferrari A, Sala P~R, Fassò A and Ranft J 2005 {\em {FLUKA: A multi-particle
  transport code (program version 2005)}\/} CERN Yellow Reports: Monographs
  (Geneva: CERN) \urlprefix\url{https://cds.cern.ch/record/898301}

\bibitem{Agostinelli:2002hh}
Agostinelli S {\em et~al.\/} (GEANT4) 2003 {\em Nucl. Instrum. Meth.\/} {\bf
  A506} 250--303

\bibitem{Paganini:2017hrr}
Paganini M, de~Oliveira L and Nachman B 2018 {\em Phys. Rev. Lett.\/} {\bf 120}
  042003 (\textit{Preprint} \eprint{1705.02355})

\bibitem{Maevskiy:2020ank}
Maevskiy A, Ratnikov F, Zinchenko A and Riabov V 2021 {\em Eur. Phys. J. C\/}
  {\bf 81} 599 (\textit{Preprint} \eprint{2012.04595})

\bibitem{Chekalina:2018hxi}
Chekalina V, Orlova E, Ratnikov F, Ulyanov D, Ustyuzhanin A and Zakharov E 2019
  {\em EPJ Web Conf.\/} {\bf 214} 02034 (\textit{Preprint} \eprint{1812.01319})

\bibitem{ATL-SOFT-PUB-2020-006}
 2020 {Fast simulation of the ATLAS calorimeter system with Generative
  Adversarial Networks} Tech. rep. CERN Geneva all figures including auxiliary
  figures are available at
  https://atlas.web.cern.ch/Atlas/GROUPS/PHYSICS/PUBNOTES/ATL-SOFT-PUB-2020-006
  \urlprefix\url{https://cds.cern.ch/record/2746032}

\bibitem{Musella:2018rdi}
Musella P and Pandolfi F 2018 {\em Comput. Softw. Big Sci.\/} {\bf 2} 8
  (\textit{Preprint} \eprint{1805.00850})

\bibitem{Fanelli:2019qaq}
Fanelli C and Pomponi J 2019 {\em Mach. Learn. Sci. Tech.\/} {\bf 1} 015010
  (\textit{Preprint} \eprint{1911.11717})

\bibitem{LHCbRICHGroup:2012mgd}
Adinolfi M {\em et~al.\/} (LHCb RICH Group) 2013 {\em Eur. Phys. J. C\/} {\bf
  73} 2431 (\textit{Preprint} \eprint{1211.6759})

\bibitem{Easo:2005xv}
Easo S, Belyaev I, Corti G, Jones C, Papanestis A, Pokorski W, Ranjard F and
  Robbe P 2005 {\em IEEE Trans. Nucl. Sci.\/} {\bf 52} 1665--1668

\bibitem{Maevskiy_2020}
Maevskiy A, Derkach D, Kazeev N, Ustyuzhanin A, Artemev M and Anderlini L 2020
  {\em Journal of Physics: Conference Series\/} {\bf 1525} 012097
  \urlprefix\url{https://doi.org/10.1088/1742-6596/1525/1/012097}

\bibitem{Forty:1998eqa}
Forty R~W and Schneider O 1998   CERN--LHCb--98--040

\bibitem{NIPS2014_5ca3e9b1}
Goodfellow I, Pouget-Abadie J, Mirza M, Xu B, Warde-Farley D, Ozair S,
  Courville A and Bengio Y 2014 {\em Advances in Neural Information Processing
  Systems\/} vol~27 ed Ghahramani Z, Welling M, Cortes C, Lawrence N and
  Weinberger K~Q (Curran Associates, Inc.)
  \urlprefix\url{https://proceedings.neurips.cc/paper/2014/file/5ca3e9b122f61f8f06494c97b1afccf3-Paper.pdf}

\bibitem{PicatosteOlloqui:2009xwb}
Picatoste~Olloqui E (LHCb) 2009 {\em J. Phys. Conf. Ser.\/} {\bf 160} 012046

\bibitem{DBLP:journals/corr/BellemareDDMLHM17}
Bellemare M~G, Danihelka I, Dabney W, Mohamed S, Lakshminarayanan B, Hoyer S
  and Munos R 2017 {\em CoRR\/} {\bf abs/1705.10743} (\textit{Preprint}
  \eprint{1705.10743}) \urlprefix\url{http://arxiv.org/abs/1705.10743}

\bibitem{arjovsky2017wasserstein}
Arjovsky M, Chintala S and Bottou L 2017 Wasserstein gan (\textit{Preprint}
  \eprint{1701.07875})

\bibitem{Aaij:2018vrk}
Aaij R {\em et~al.\/} 2019 {\em EPJ Tech. Instrum.\/} {\bf 6} 1
  (\textit{Preprint} \eprint{1803.00824})

\bibitem{sweight2005}
Pivk M and Le~Diberder F 2005 {\em Nuclear Instruments and Methods in Physics
  Research Section A: Accelerators, Spectrometers, Detectors and Associated
  Equipment\/} {\bf 555} 356–369 ISSN 0168-9002
  \urlprefix\url{http://dx.doi.org/10.1016/j.nima.2005.08.106}

\bibitem{sweight2019}
Borisyak M and Kazeev N 2019 {\em Journal of Instrumentation\/} {\bf 14}
  P08020–P08020 ISSN 1748-0221
  \urlprefix\url{http://dx.doi.org/10.1088/1748-0221/14/08/P08020}

\bibitem{Kostenetskiy_2021}
Kostenetskiy P~S, Chulkevich R~A and Kozyrev V~I 2021 {\em Journal of Physics:
  Conference Series\/} {\bf 1740} 012050
  \urlprefix\url{https://doi.org/10.1088/1742-6596/1740/1/012050}

\end{thebibliography}
\end{document}